\documentclass[a4paper]{jpconf}
\usepackage{graphicx}

\def\farcs{\hbox{$.\!\!^{\prime\prime}$}}

\begin{document}
\title{Mid--infrared interferometry of massive young stellar objects}

\author{H Linz$^1$, B Stecklum$^2$, R Follert$^{1,2}$, Th Henning$^1$, 
        R van Boekel$^1$, A Men'shchikov$^3$, I Pascucci$^4$ and M Feldt$^1$}

\address{$^1$ MPIA Heidelberg, K\"onigstuhl 17, D--69117 Heidelberg, Germany\\
         $^2$ Th\"uringer Landessternwarte Tautenburg, Sternwarte 5, D--07778
	 Tautenburg, Germany \\
	 $^3$ CEA, IRFU, Service d'Astrophysique, F--91191 Gif-sur-Yvette,
	 France \\
	 $^4$ Steward Observatory, Department of Astronomy, University of 
	      Arizona, 933 N Cherry Ave., Tucson AZ 85721-0065, U.S.A.}

\ead{[linz,follert,henning,boekel,mfeldt]@mpia.de, stecklum@tls-tautenburg.de,
      alexander.menshchikov@cea.fr, pascucci@as.arizona.edu}

\begin{abstract}
The very inner structure of massive YSOs is difficult to trace. 
With conventional observational methods we identify structures still several hundreds
of AU in size. However, the (proto-)stellar growth takes place at the innermost regions 
($<$100 AU) where the actual mass transfer onto the forming high-mass star occurs. 
We present results from our programme toward massive YSOs at the VLTI, utilising the 
two-element interferometer MIDI. To date, we observed 10 well-known massive YSOs down to 
scales of 20 mas (typically corresponding to 20 - 40 AU for our targets) in the 8-13 
micron region. We clearly resolve these objects which results in low visibilities and 
sizes in the order of 30-50 mas. For two objects, we show results of our modelling. 
We demonstrate that the MIDI data can reveal decisive structure information for massive 
YSOs. They are often pivotal in order to resolve ambiguities still immanent in model 
parameters derived from sole SED fitting. 

\end{abstract}

\section{Introduction}
High-mass stars predominantly form in clustered environments much  
more distant than typical well-investigated low-mass star-forming regions. 
Thus, high spatial resolution is a prerequisite for making progress in 
the observational study of high-mass star formation. Furthermore, all 
pre-main sequence phases are usually deeply embedded.  This often 
forces observers of embedded massive young stellar objects (MYSOs) 
to move to the mid-infrared (MIR) where the resolution of conventional 
imaging is limited to $>$ 0\farcs25 even with 8-m class telescopes. Hence, 
one traces linear scales still several hundred AU in size even for the  
nearest MYSOs, and conclusions on the geometry of the innermost  
circumstellar material remain ambiguous. MIR emission moderately  
resolved with single telescopes may even arise from the inner  
outflow cones \cite{2006ApJ...642L..57D,2005A&A...429..903L}. \\  
A versatile method to overcome the diffraction limit of single  
telescopes is to employ interferometric techniques. We are  
conducting a larger survey toward MYSOs based on MIR interferometry. In  
total, we have observed 10 sources so far.  All these sources, mostly  
comprising BN-type objects \cite{1990FCPh...14..321H}, are clearly  
resolved with the interferometer baselines we applied ($\ge 16$ m). This  
in itself is a major step forward compared to the previous more or  
less unresolved thermal infrared imaging with 4- to 8-m class  
telescopes for these sources.

\section{Observations and Modelling}\label{Sec:Modelling}

Visibilities in the mid-infrared wavelength range 8--13\,$\mu$m
have been obtained with the instrument MIDI 
\cite{2003SPIE.4838..893L} at the Very Large Telescope Interferometer.
Within the framework of Guaranteed Time Observations for MIDI as well as 
Director's Discretionary Time, we observed our objects at several 8.2-m
Unit Telescope (UT) baseline length / baseline orientation combinations 
between June 2004 and May 2006. Additionally, first  observations with
the 1.8-m Auxiliary Telescopes (ATs) have been performed in autumn 2006
and spring 2007 for three of our sources.  \\
We refer to \cite{2004A&A...423..537L} for a more detailed 
description of the standard observing procedure for MIDI observations. 
For all our observations, the so-called HighSens mode was used:
during self-fringe tracking, all the incoming thermal infrared 
signal is used for beam combination and fringe tracking, while the 
photometry is subsequently obtained in separate observations. We use the 
MIDI prism as the dispersing element, hence, we finally get spectrally 
dispersed visibilities with a spectral resolution of $R  \approx  30$.
HD 169916 was used as the main interferometric and photometric 
standard star (Procyon in the case of the Orion BN object) and was observed 
always immediately after the science objects. 
In addition, all calibrator measurements of a night were collected to 
create an average interferometric transfer function and to assign error 
margins to the measured visibilities. 
We have reduced the interferometric data with the MIA+EWS package, version 
1.5, developed at the MPIA Heidelberg and the University of Leiden. \\
One way to interpret visibilities is to use a combination of simple geometrical
configurations like Gaussians, disks, rings, or point sources to construct 
a synthetic intensity distribution similar to the observed one.
Still, to learn more about the potentially more complicated intensity structure,
such an ansatz might not always be sufficient. Further, more physically 
motivated modelling is often necessary for interpretation. \\
We apply self-consistent continuum radiative transfer modelling in 
order to produce synthetic MIR intensity maps and to compare their spatial 
frequency spectrum with the observed visibilities. Here, we are mainly concerned
with the question which spatial distribution of the circumstellar material can 
account for both, the SED and the visibilities of our targets. \\
We used the SED online fitting tool of Robitaille et al.~\cite{2007ApJS..169..328R}  
that can in principle comprise YSO models including 
an envelope plus circumstellar disk. We refer to this publication for details
on the setup of these models. For parameter combinations well fitting the SED, 
the underlying radiative transfer code of Whitney et al.~\cite{2003ApJ...591.1049W} 
is then used to produce high-resolution MIR 
intensity maps for these selected models. Comparing the resulting synthetic
visibilities with the observed ones indicates which models finally account for 
the observed SED {\it and} visibilities.

\section{Results}

In this contribution, we concentrate on two objects for which we have reached
considerable progress. They show visibilities in the order of 0.2\,--\,0.3 at 
UT baselines.
We note that such visibilities, although not reaching the relatively high 
levels of most Herbig Ae/Be stars \cite{2004A&A...423..537L}, are 
qualitatively different from the very low visibilities (0.01\,--\,0.05) found 
for several of the other objects in our sample as well as recently reported 
for two other massive YSOs \cite{2007ApJ...671L.169D,2008ASPC..387..444V}.

\subsection{M8E--IR: A BN-type object with a bloated central 
            star?}\label{Sec:M8E-IR}  

This is a prominent BN-type MYSO at a distance  
of roughly 1.5 kpc.  Although M8E-IR was a well  
investigated object in the 1980's, the spatial resolution for most of  
the IR observations of M8E-IR was poor. An exception is the work by  
\cite{1985ApJ...298..328S} who speculated on the existence of a small  
circumstellar disk around M8E-IR based on thermal infrared  
lunar occultation data. 
\begin{figure}[ht]  
\hspace*{-0.0cm}\includegraphics[width=8.5cm]{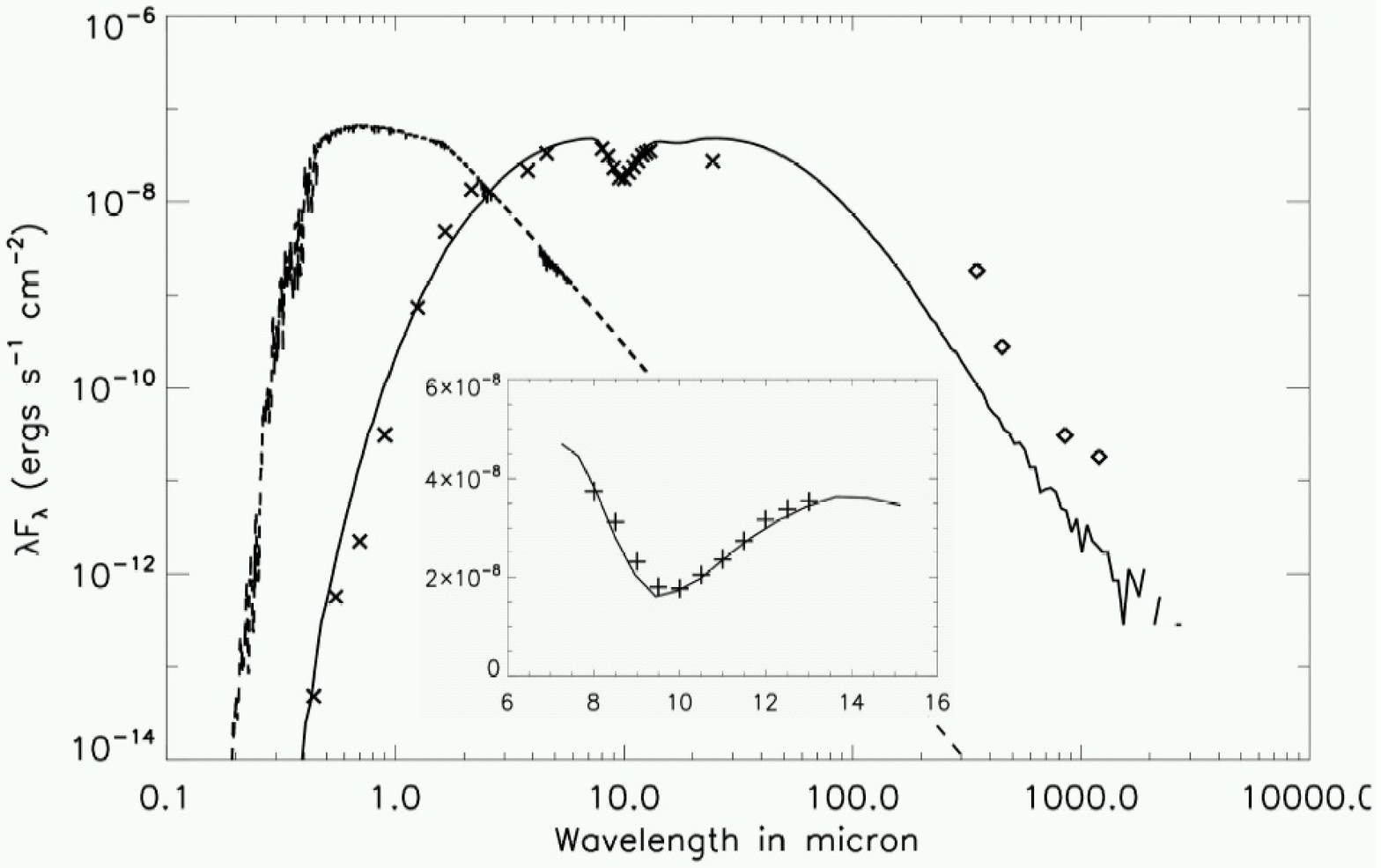}\hspace*{0.3cm}\vspace*{0.25cm}\includegraphics[width=7.5cm]{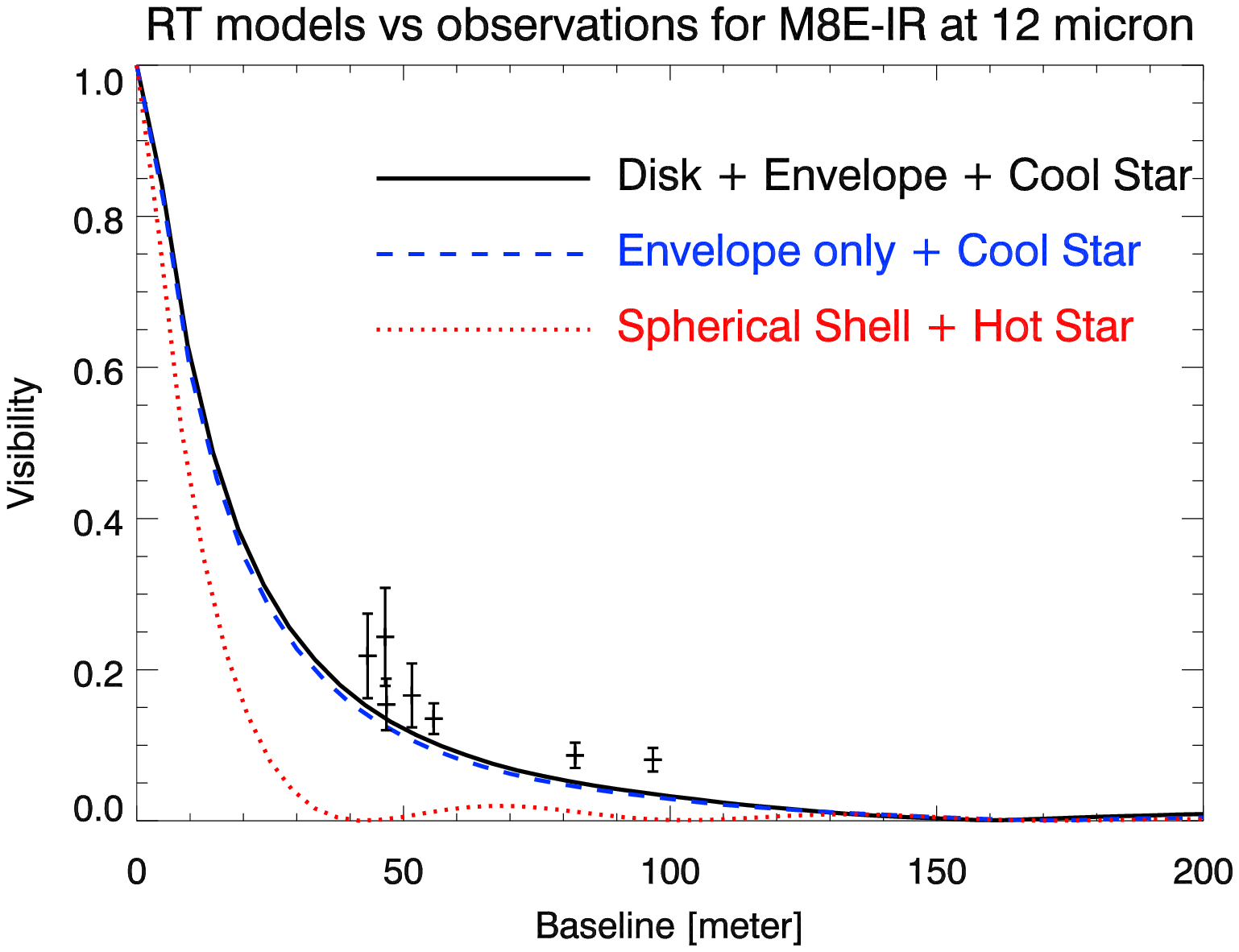}  
\caption{{\bf Left:} The SED of M8E-IR, shown as crosses (measured fluxes) and diamonds (upper limits). The solid curve denotes the
SED of the best-fitting radiative transfer model. The dashed line marks the unreddened SED of the bloated central star for this particular 
model. The inset is a zoom into the 8\,--\,13 micron region that underlines the quality of the fit.  
{\bf Right:} Comparison of the best-fitting models with bloated cool central star (black solid line and blue dashed line) and a standard 
configuration including a spherically symmetric shell and a hot central star (red dotted line). Obviously, the latter one is not 
compatible with the observed visibility data, shown here as plus signs including error bars.}\label{Fig:M8EIR}  
\end{figure}  

\noindent
For SED fitting, we use the M8E-IR photometric data collected in   
\cite{2002aApJS..143..469M} plus new 1.2 mm data from \cite{2006A&A...447..221B}.   
We want to stress that no (sub-)millimeter interferometry   
on M8E-IR is reported in the literature which could spatially disentangle   
the flux contributions from M8E-IR and the radio source 8$''$ away. Hence, we   
consider the M8E-IR fluxes for $\lambda > 24.5 \, \mu$m just as upper limits in   
our modelling. Furthermore, we include new optical photometry in the B, V, and   
I filters reported by \cite{2005A&A...430..941P} as well as our new Subaru
24.5 $\mu$m photometry. In addition, the 8--13 $\mu$m total  
flux spectrum taken in the course of the MIDI measurements was used to further 
constrain the multitude of viable models.  
We show the SED of the best fitting model in Fig.~\ref{Fig:M8EIR} (Left). \\
The best-fit model comprises a very compact circumstellar disk 
($<$ 50  AU), a larger envelope with small bipolar cavities, and a cool central object
($T_{\rm eff} \sim $ 4500 K). 
We mention explicitly, that  among the well-fitting models there are also 
configurations without a disk  (axisymmetric flattened envelope + outflow cavities only), 
but also including a cool bloated central star. These nevertheless give almost the 
same high visibilities. This suggests that in the case of M8E--IR, the choice of 
the central object might actually govern the resulting visibility levels (see below).\\  
Traditionally, such BN-type objects have been modelled as a spherical dust 
shell surrounding a hot, non--bloated central star \cite{1990FCPh...14..321H}.
To check if even such canonical configurations can account for the SED and the 
visibilities we employed well-tested models, based on the code used in 
\cite{1999ApJ...519..257M}. With a purely spherically symmetric geometry 
and a 24,000 K central star it is possible to find models that reasonably fit 
the SED of M8E-IR. 
Still, compared to the measured visibilities, these models result in far too 
low visibilities ($<$ 0.05) for M8E-IR in the baseline range 30--60 m 
over the whole 8--13 $\mu$m range. The error margins of the MIDI visibilities 
(on average 10\%) do not account for such large differences.
In Fig.~\ref{Fig:M8EIR} (Right), as an example the {\it  u,v}-spectrum of the 12 $\mu$m 
synthetic images based on the cool star models are included as black solid and 
blue dashed line. They are compared to the corresponding image  
from the above-mentioned spherically symmetric modelling with hot central star 
(red dotted  line).  Although the {\it u,v}-spectra of the models with cool 
central object still show somewhat lower visibilities than the measured ones, 
obviously they are qualitatively different from the spherically symmetric model
with hot central star which fails to match the observational data.\\
The best-fitting models for M8E-IR in the Robitaille  
model grid feature central stars of 10--15 M$_\odot$ which are  
strongly bloated (120\,--\,150 R$_\odot$) and, therefore, have  
relatively low effective temperatures.  Such solutions can occur since  
the Robitaille grid comprises the full range of canonical  
pre-main sequence evolutionary tracks from the Geneva  
group 
as possible parametrisation of the central objects. M8E-IR probably  
cannot straightforwardly be identified with correspondingly very  
early evolutionary stages, and we refer to \cite{2008ASPC..387..290R} 
for an extensive discussion on the intricate dependencies in the model grid. 
However, the tendency for a  bloated central star in the case of M8E-IR may 
be valid. As demonstrated already in \cite{1977A&A....54..539K}, accretion with 
high rates onto  main sequence stars can temporarily puff up such stars. 
Further encouragement to consider this comes from recent modelling of the 
pre-main-sequence evolution of stars in dependence of the accretion rate 
\cite{Hosokawa2, 2008ASPC..387..189Y}. These groups find that for accretion 
rates (onto the forming star) reaching 10$^{-3} \, $M$_\odot$/yr, the protostellar  
radius can temporarily increase to $> 100$\,R$_\odot$, in accordance  
with our indirect findings from the model fitting.  Interestingly,  
\cite{1988ApJ...327L..17M} revealed high-velocity molecular outflows  
from M8E-IR based on M-band CO absorption spectra and speculated on  
recent ($<120$ yr) FU Ori-type outbursts for this object. If these  
multiple outflow components really trace recent strong accretion  
events, the central star could have indeed been affected.  
In addition, M8E-IR is not detected by cm observations with medium  
sensitivity \cite{1984ApJ...278..170S, 1998A&A...336..339M}. This  
could be explained by large accretion rates still quenching a forming  
hypercompact H{\sc ii} region \cite{1995aRMxAC...1..137W}.  
Furthermore, also a bloated central star with $T_{\rm eff} \ll 10000$ K  
would give a natural explanation for these findings.

\subsection{The KW Object in M17: An example for the transition towards
            the Herbig Be star stage}

\begin{figure}[t]  
\hspace*{-0.0cm}\includegraphics[width=8.5cm]{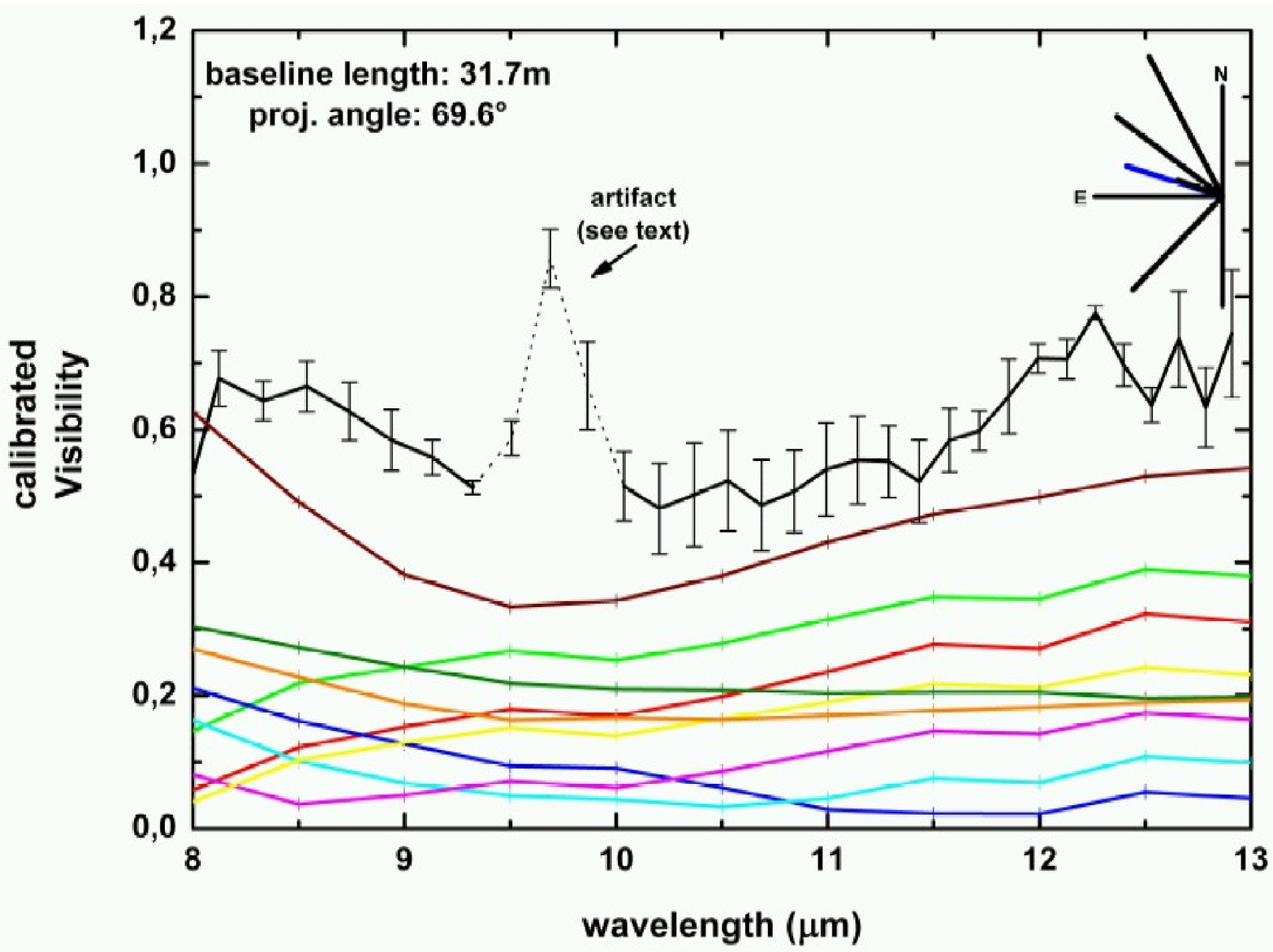}\hspace*{0.5cm}\includegraphics[width=6.5cm]{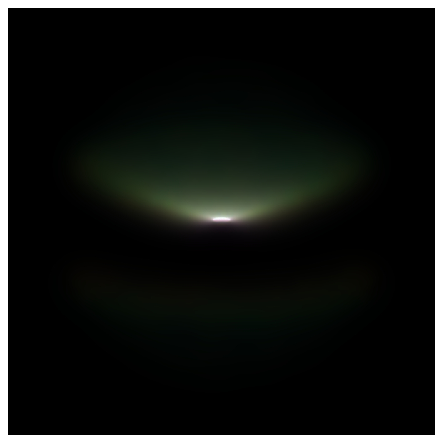}  
\caption{{\bf Left:} Observed visibility curve (black line with error bars) for the 32-m AT baseline for the KW Object in M\,17. The coloured lines
give cuts through the synthetic MIR maps of different models well fitting the SED. The 75$^\circ$ inclination model (brown curve) clearly 
stands out as coming closest to the observed visibilities.
{\bf Right:} Synthetic 8\,--\,13 $\mu$m image (in logarithmic stretch) for the 75$^\circ$ inclination model delivered by the Robitaille 
fitter for the KW Object (see left panel). Displayed is a linear size of 1400 $\times$ 1400 AU$^2$. This model has a disk 
radius of 500 AU and a 15 M$_\odot$  central star with $T_{\rm eff}$ = 31000 K. As an important feature, the inner bright disk rim
is directly visible which leads to higher visibilities than for most edge-on configurations.}\label{Fig:KWO}  
\end{figure}

In order to give another example for the abilities of MIR  
interferometry to assess the viability of different geometric models  
for a massive YSO, we report here on results for the  
object M\,17 IRS1 \cite{Chini}, also called the Kleinman-Wright (KW) Object.  
We have obtained three visibility measurements with different position  
angles for UT baselines from 43~m to 56~m. Furthermore, for the first time, we
show here MIDI observations towards massive YSOs utilising the 1.8-m ATs.
We find clearly different  visibility levels for the five measurements, 
staying at a 0.1--0.3 level  for the UT measurement but reaching more than 0.5 
for the shorter AT baselines. Also here, we consulted the Robitaille SED fitter 
in order to find  models reproducing the SED. Since the SED of M\,17 IRS1 is less  
constrained in the literature than in the case of M8E-IR, the fitter  
allows for a larger variety of models, all however including an  
intermediate-sized circumstellar disk and only a very small envelope 
contribution. In Fig.~\ref{Fig:KWO} (Left) we show one example where the 32-m AT
measurement is compared to 9 well-fitting models. Here, MIDI can be used as a
discriminator. Among the SED-fitting models it rejects very strongly inclined (87$^\circ$)  
edge-on models where we see relatively diffuse MIR emission only,  
since the view onto the central compact disk rim is obstructed by the outer disk.  Consequently,  
such models result in far too low N-band visibilities and do  
{\it not} reach the measured elevated visibility levels for any  
baseline orientation. The less inclined 75$^\circ$ model to the right  
at least provides a direct view onto the inner hot disk rim and  
therefore results in clearly higher visibilities than edge-on  
models.  

\subsection{Beyond visibilities: Differential phases}

\begin{figure}[ht]  
\centering  
\hspace*{-0.0cm}\includegraphics[width=8.5cm]{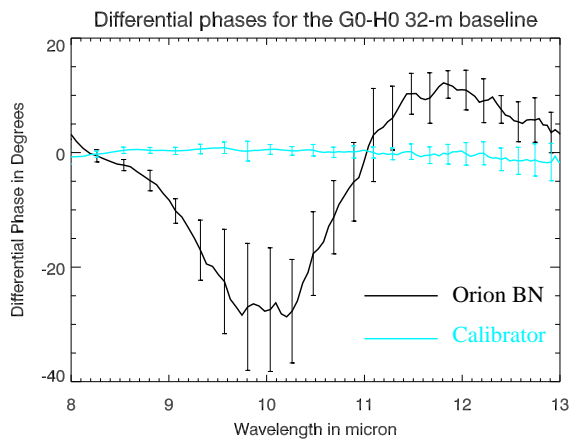}  
\caption{Differential phases for the BN object in Orion. Shown are the data
from the 32-m AT baseline, including the 3-$\sigma$ error bars, for BN and the
calibrator closest in time to the science object. Under ideal observing
conditions, the differential phases for a calibrator ought to be constant at 
zero degrees. The significant differential phase signal for BN indicates
a deviation from spherical symmetry on scales of around 
75 milli-arcseconds traced by this baseline.}\label{Fig:BN-phase}  
\end{figure}

For two-element infrared interferometers like MIDI, the {\it visibility} (and
the correlated flux, respectively) is the typical quantity from which further
information is derived. To calibrate the absolute position of the fringe phases directly is 
very difficult in this wavelength regime, and for closure phase measurements at least three beams 
are necessary. However, it is possible to extract the {\it differential phase} for the MIDI data
within the coherent data reduction algorithm presented in
\cite{2004SPIE.5491..715J}. This quantity can tell us about the geometry 
of the object under investigation. Point-symmetric objects have a differential phase 
of zero. Differential phases deviating from this are related to position shifts of the 
photo-centre over wavelength. This can be considered as an analogon
to the signal derived from conventional spectro-astrometry measurements. \\
In Fig.~\ref{Fig:BN-phase} we show the differential phases for the well-known BN object 
in the Orion BN/KL region which we recently observed at several short MIDI baselines with 
the Auxiliary Telescopes. A significant differential phase signature is obvious. This 
supports recent claims that Orion BN is deviating from spherical symmetry, reported in 
\cite{2005Natur.437..112J}, where these authors infer the presence of a compact circumstellar disk 
by means of near-IR polarimetric observations.

\section{Conclusion}    
We have observed massive young stellar objects with the  
MIR interferometer MIDI at the VLTI. We find substructures with MIR sizes   
around 30\,--\,50~mas. Using the measured visibilities as discriminator, we can exclude  
purely spherically symmetric matter distributions with a hot central star for   
the object M8E-IR. The most probably configuration consists of a compact   
circumstellar disk surrounded by a larger envelope plus a cool central object. The total disk size   
is not well constrained by the models, and the data allow also for an envelope-only
configuration. The finding of a cool central object inside this MYSO is consistent with the idea
that the 10--15 M$_\odot$ central star has been strongly bloated by recent strong accretion events. \\  
For M\,17 IRS1, we can exclude the nearly edge-on disk models which were among the models suggested 
by the SED fitting. Instead, a moderately inclined circumstellar disk might be  
closer to the truth, and the results of a disk-dominated object with small envelope contribution support 
an earlier suggestion that M\,17 IRS1 is near to the Herbig Be star phase.\\  
Our results show that IR interferometry is a viable tool to reveal decisive 
structure information on embedded MYSOs and to resolve ambiguities arising from 
fitting the spectral energy distribution. With the inclusion of the Auxiliary Telescopes,
the VLTI is currently becoming even more flexible to tackle such tasks. Finally, 
the 2$^{\rm nd}$-generation VLTI instrument {\sc Matisse} \cite{2006SPIE.6268E..31L} for the 
thermal IR will reveal even more complex details of MYSOs by adding closure phase and 
imaging capabilities to MIR interferometry.

\end{document}